\documentclass[twocolumn]{svjour3}          

\usepackage{cite,graphicx,pstool,psfrag}
\usepackage{mathrsfs,amsmath,amsfonts,amssymb}
\usepackage[nolist]{acronym} 
\usepackage[multiple]{footmisc}
\usepackage{bm,wrapfig}
  \usepackage{flushend}

\newcommand{\N}[2]{\mathcal{N}\!\left(#1,#2\right)}       		
\newcommand{\EX}[1]{{\mathbb{E}}\left\{{#1}\right\}}        		

\def\Hz{\mathcal{H}_{0}}                               	
\def\Ho{\mathcal{H}_{1}}                                    
\def\Hj{\mathcal{H}_{j}}                                    	

\def\testHoneOverHzeroL{\begin{array}{c}
		{\mathcal{H}}_1 \vspace{-0.045cm}\\
		\gtrless \vspace{-0.02cm}\\
		{\mathcal{H}}_0
	\end{array}}

\def\testHzeroOverHoneL{\begin{array}{c}
    {\mathcal{H}}_0 \vspace{-0.045cm}\\
    \gtrless \vspace{-0.02cm}\\
    {\mathcal{H}}_1
  \end{array}}
\def\Pd{\textrm{P}_\textrm{D}}

\def\Pfa{\textrm{P}_\textrm{FA}}

\def\Hz{\mathcal{H}_\textrm{0}}

\newcommand{\hmt}[1]{{#1}^{\textrm{H}}}  		
\newcommand{\transp}[1]{{#1}^{\textrm{T}}}               
\newcommand{\tr}[1]{\textrm{tr}\!\left\{#1\right\}}         	

\newcommand{\diag}[1]{\textrm{diag}\!\left\{#1\right\}}                				
\newcommand{\Id}[1]{{\mathbf I}_{#1}}                                        		

\def\SNR{\textsf{SNR}}

\def\SNRref{\textsf{SNR}_\textrm{ref}}
\def\sig{\sigma^2}

\def\sigt{\sigma_\textrm{t}^2}
\def\sigs{\sigma^2_{\text{s}}}

\def\sigestHz{\widehat{\sigma}^2_{\Hz}}
\def\sigestHo{\widehat{\sigma}^2_{\Ho}}
\def\sigt{\sigma^2_{\text{t}}}

\def\N{N}
\def\yvk{\mathbf{y}_{k}}

\def\yvi{\mathbf{y}_{i}}

\def\ytildevk{\widetilde{\mathbf{y}}_{k}}

\newcommand{\yv}[1]{\mathbf{y}_{#1}}

\newcommand{\nv}[1]{\mathbf{n}_{#1}}
\newcommand{\ntildev}[1]{\widetilde{\mathbf{n}}_{#1}}
\newcommand{\sv}[1]{\mathbf{s}_{#1}}

\def\Y{\mathbf{Y}}                      

\def\Sig{\mathbf{\Sigma}}
\def\Sigz{\widetilde{\mathbf{\Sigma}}_{0}}
\def\Sigo{\widetilde{\mathbf{\Sigma}}_{1}}
\def\Sigj{\widetilde{\mathbf{\Sigma}}_{j}}

\def\SCM{\mathbf{S}}                    
\def\R{\mathbf{R}}                      
\def\W{\mathbf{W}}

\def\T{\text{T}}

\def\Tind{\T^{\left(\text{ind}\right)}} 
\def\Tsph{\T^{\left(\text{sph}\right)}} 

\def\lambdai{\lambda_{i}}
\def\lambdaBi{\lambda^{\B}_{i}}

\def\lmax{\lambda_{1}}
\def\mui{\mu_{i}}

\def\muKi{\mu^{\K}_{i}}

\def\fs{f_{\textrm{s}}}

\def\K{K}
	
\def\n{{n}}

\def\p{{p}}

\def\Tfc{\T_{\text{fc}}}
\def\TedA{\T_{\text{ed-all}}}
\def\Ted{\T_{\text{ed}}}
\def\Tenped{\T_{\text{enp-ed}}}
\def\Twenped{\T_{\text{w-enp-ed}}}

\def\Tfagm{\T_{\text{f-agm}}}

\def\Tmos{\T_{\text{mos}}}
\def\TmosB{\T_{\text{search-B}}}
\def\Tsf{\T_{\text{sf}}}
\def\Twsf{\T_{\text{sf}}}
\def\Twfagm{\T_{\text{wf-agm}}}
\def\Tfkl{\T_{\text{fkl}}}

\def\Tlbi{\T_{\text{lbi}}}
\def\Tsph{\T_{\text{sph}}}
\def\Tmet{\T_{\text{met}}}
\def\Tmme{\T_{\text{mme}}}
\def\Tind{\T_{\text{ind}}}

\def\Tfro{\T_{\text{fro}}}
\def\Tmax{\T_{\text{max}}}

\def\Twlbi{\T_{\text{w-lbi}}}
\def\Twsph{\T_{\text{w-sph}}}
\def\Twind{\T_{\text{w-ind}}}
\def\Twmax{\T_{\text{w-max}}}

\def\Tac{\T_{\text{ac}}}
\def\Tacone{\T_{\text{ac-1}}}

\def\Tcov{\T_{\text{cov}}}

\def\Nfft{{N}_\text{fft}}

\def\Nsamp{{N}_\text{avg}}

\def\P{\bm{p}}
\def\W{\bm{w}}

\def\Pi{P_{i}}

\newcommand{\norm}[2]{||#1||_{#2}}

\def\SCMz{\mathbf{S}_\text{0}}
\def\B{\mathbf{B}}
\def\K{\mathbf{K}}
\def\Rz{\mathbf{R}_\text{0}}
\def\OSF{\mathsf{OSF}}

\def\mui{\mu_{i}}

\def\Sset{\mathcal{S}}
\def\Ssetbar{\bar{\mathcal{S}}}
\def\SCMf{\widetilde{\mathbf{S}}}
\def\stildeii{\widetilde{s}_{i,i}}
\def\fL{f_{\text{L}}}
\def\fH{f_{\text{H}}}
\def\Band{\mathcal{B}}
\def\Bandbar{\overline{\mathcal{B}}}

\def\PB{\P_{\left[\Band\right]}}
\def\QB{{\bm q}_{\left[\Band\right]}}
\def\PfsB{\PBbar}
\def\QfsB{\QBbar}
\def\PBbar{\P_{[\Bandbar]}}
\def\QBbar{{\bm q}_{[\Bandbar]}}

\def\BandCB{\mathcal{C}_{B}}
\def\PCB{\P_{\left[\BandCB\right]}}

\begin{acronym}
\scriptsize
\acro{ADC}{analog to digital converter}
\acro{AWGN}{additive white Gaussian noise}
\acro{WGN}{white Gaussian noise}
\acro{CDF}{cumulative distribution function}
\acro{c.d.f.}{cumulative distribution function}
\acro{CR}{cognitive radio}
\acro{CV}{coefficient of variation}
\acro{DAA}{detect and avoid}
\acro{DFT}{discrete Fourier transform}
\acro{DVB-T}{digital video broadcasting\,--\,terrestrial}
\acro{ECC}{European Community Commission}
\acro{FCC}{Federal Communications Commission}
\acro{FFT}{fast Fourier transform}
\acro{GPS}{Global Positioning System}
\acro{i.i.d.}{independent, identically distributed}
\acro{IFFT}{inverse fast Fourier transform}
\acro{ITU}{International Telecommunication Union}
\acro{MB}{multiband}
\acro{MC}{multicarrier}
\acro{MF}{matched filter}
\acro{MI}{mutual information}
\acro{MIMO}{multiple-input multiple-output}
\acro{MRC}{maximal ratio combining}
\acro{MMSE}{minimum mean-square error}
\acro{OFDM}{orthogonal frequency-division multiplexing}
\acro{p.d.f.}{probability density function}
\acro{PAM}{pulse amplitude modulation}
\acro{PSD}{power spectral density}
\acro{PSK}{phase shift keying}
\acro{QAM}{quadrature amplitude modulation}
\acro{QPSK}{quadrature phase shift keying}
\acro{r.v.}{random variable}
\acro{R.V.}{random vector}
\acro{SNR}{signal-to-noise ratio}
\acro{SS}{spread spectrum}
\acro{TH}{time-hopping}
\acro{ToA}{time-of-arrival}
\acro{UWB}{ultrawide band}
\acro{UWBcap}[UWB]{Ultrawide band}
\acro{WLAN}{wireless local area network}
\acro{WMAN}{wireless metropolitan area network}
\acro{WPAN}{wireless personal area network}
\acro{WSN}{wireless sensor network}

\acro{ML}{maximum likelihood}
\acro{GLR}{generalized likelihood ratio}
\acro{GLRT}{generalized likelihood ratio test}
\acro{LLRT}{log-likelihood ratio test}
\acro{LRT}{likelihood ratio test}
\acro{$P_{EM}$}{probability of emulation, or false alarm}
\acro{$P_{MD}$}{probability of missed detection}
\acro{$P_D$}{probability of detection}
\acro{$P_{FA}$}{probability of false alarm}
\acro{ROC}{receiver operating characteristic}

\acro{AGM}{arithmetic-geometric mean ratio test}
\acro{AIC}{Akaike information criterion}
\acro{BIC}{Bayesian information criterion}
\acro{BWB}{bounded worse beaviour}
\acro{CA-CFAR}{cell-averaging constant false alarm rate}
\acro{CAIC}{consistent AIC}
\acro{CAICF}{consistent AIC with Fisher information}
\acro{CFAR}{constant false alarm rate}
\acro{CDR}{constant detection rate}
\acro{CP}{cyclic prefix}
\acro{CPC}{cognitive pilot channel}
\acro{CL}{centroid localization}
\acro{CSCG}{circularly-symmetric complex Gaussian}
\acro{d.o.f.}{degrees of freedom}
\acro{DC}{distance based combining}
\acro{DSM}{dynamic spectrum management}
\acro{DSA}{dynamic spectrum access}
\acro{DSP}{digital signal processing}
\acro{DTV}{digital television}
\acro{ECC}{European Communications Committee}
\acro{ED}{energy detector}
\acro{EDC}{efficient detection criterion}
\acro{EGC}{equal gain combining}
\acro{EME}{energy to minimum eigenvalue ratio}
\acro{ENP}{estimated noise power}
\acro{ERC}{European Research Council}
\acro{ES-WCL}{energy based selection WCL}
\acro{EU}{European Union}
\acro{ExAIC}{exact AIC}
\acro{ExBIC}{exact BIC}
\acro{FC}{fusion center}
\acro{FIM}{Fisher information matrix}
\acro{GLM}{general linear model}
\acro{GLRT}{generalized likelihood ratio test}
\acro{GIC}{generalized information criterion}
\acro{HC}{hard combining}
\acro{HF}{hard fusion}
\acro{IoT}{internet-of-things}
\acro{ITC}{information theoretic criteria}
\acro{K-L}{Kullback-Leibler}
\acro{LBI}{locally best invariant}
\acro{LR}{likelihood ratio}
\acro{LS}{least square}
\acro{MDL}{minimum description length}
\acro{MET}{ratio of maximum eigenvalue to the trace}
\acro{ML}{maximum likelihood}
\acro{MOS}{model order selection}
\acro{MTM}{multi taper method}
\acro{MME}{maximum to minimum eigenvalues ratio}
\acro{NP}{Neyman-Pearson}
\acro{OFCOM}{Office of Communications}
\acro{OSA}{opportunistic spectrum access}
\acro{OSF}{oversampling factor}
\acro{PMSE}{programme making and special events}
\acro{PU}{primary user}
\acro{PWCL}{power based WCL}
\acro{r.o.f.}{roll-off factor}
\acro{RAN}{Regional Area Network}
\acro{RF}{radio-frequency}
\acro{RMSE}{root mean square error}
\acro{RWCL}{relative WCL}
\acro{RSS}{received signal strength}
\acro{SC}{soft combining}
\acro{SCM}{sample covariance matrix}
\acro{SCovM}{sample covariance matrix}
\acro{SCorM}{sample correlation matrix}
\acro{SDR}{software-defined radio}
\acro{SF}{soft fusion}
\acro{SP}{sensing period}
\acro{SS}{spectrum sensing}
\acro{SSE}{signal subspace eigenvalues}
\acro{StD}{signal-to-be-detected}
\acro{SU}{secondary user}
\acro{TOA}{time of arrival}
\acro{TDOA}{time difference of arrival}
\acro{TVBD}{TV Bands Device}
\acro{TS-WCL}{two step WCL}
\acro{USRP}{universal software radio platform}
\acro{WCL}{weighted centroid localization}
\acro{WSD}{white space device}
\acro{WSN}{wireless sensor network}
\acro{WSS}{wideband spectrum sensing}

\end{acronym}

\begin{document}

\title{On Oversampling-Based Signal Detection\thanks{This work was supported in part by MIUR under the program ``Dipartimenti di Eccellenza (2018-2022) - Precise-CPS,'' and in part by the EU project eCircular (EIT Climate-KIC). The material in this paper was presented in part at the IEEE Int. Symp. on Personal, Indoor and Mobile Radio Comm. (PIMRC 2018), Bologna, Italy, Sep. 2018.}}
\subtitle{A pragmatic approach}
%
%
\author{Andrea~Mariani \and Andrea~Giorgetti \and  Marco~Chiani}
%
%
\institute{A.~Mariani \at
              Onit Group S.r.l. \\
              Via dell'Arrigoni 308, 47522 Cesena (FC), Italy\\
              \email{amariani@onit.it}           
              \and
              A.~Giorgetti and M.~Chiani \at
              CNIT, IEIIT/CNR, Dept. of Electrical, Electronic, and Information Engineering ``Guglielmo Marconi'' (DEI)\\ University of Bologna\\
		      Via dell'Universit\`a 50, 47522 Cesena (FC), Italy\\
		      \email{andrea.giorgetti@unibo.it}\\
              \email{marco.chiani@unibo.it}
}
\date{Received: 31 January 2019 / Revised: 14 June 2019 / Accepted: 6 July 2019}

\maketitle

\begin{abstract}
The availability of inexpensive devices allows nowadays to implement \ac{CR} functionalities in large-scale networks such as the \ac{IoT} and future mobile cellular systems. In this paper, we focus on \ac{WSS} in the presence of oversampling, i.e., the sampling frequency of a digital receiver is larger than the signal bandwidth, where signal detection must take into account the front-end impairments of low-cost devices. Based on the noise model of a \ac{SDR} dongle, we address the problem of robust signal detection in the presence of noise power uncertainty and non-flat noise \ac{PSD}. In particular, we analyze the \ac{ROC} of several detectors in the presence of such front-end impairments, to assess the performance attainable in a real-world scenario. We propose new frequency-domain detectors, some of which are proven to outperform previously proposed spectrum sensing techniques such as, e.g., eigenvalue-based tests. The study shows that the best performance is provided by a noise-uncertainty immune \ac{ED} and, for the colored noise case, by tests that match the \ac{PSD} of the receiver noise.
\end{abstract}

\acresetall  
\section{Introduction}\label{sec:introduction}

The increasing demand for a limited resource such as the \ac{RF} spectrum is the propelling force toward new ways radio ecosystems can coexist. 
For example, the coexistence between radar and wireless communications is a topic which recently received increasing attention by the DARPA and the US National Spectrum Consortium \cite{nsc,ChiGioPao:J18}.
To this aim, \ac{CR} systems represent a paradigm to opportunistically access the spectrum provided that the \ac{RF} environment is monitored through the so-called \ac{SS} \cite{KanGio:12,ShaMarGioMitChi:15}. The objective of \ac{SS} is to infer the presence of signals in a frequency band, thus identifying spectrum holes and enabling \ac{PU} protection. 

Software-defined radio (\acsu{SDR}) was born with the aim of building flexible front-ends for transceivers in which radio functionalities are controlled and programmable by software \cite{Mit:93,Bur:00,Jon:05,Wyg:16}. In the last decade, \ac{SDR} gained a renewed interest for implementing multi-band multi-standard platforms, in particular to enable \ac{CR} \cite{Bur:00,Jon:05,Bag:06,Abi:07,Wyg:16,MatlabSDR}.
In parallel, we have seen the development and diffusion of several general purpose high-performance \acp{SDR} such as \acp{USRP} \cite{usrp}.   
However, in the context of \ac{IoT} low-cost low-complexity \acp{SDR} appear more appropriate. There are several examples of such \acp{SDR} that can be used for \ac{SS} and, among them, one very popular is represented by the \ac{DVB-T} dongle based on the Realtek RTL2832U chipset.

The adoption of low-cost devices is, on the other hand, critical for \ac{WSS} due to the presence of receiver nonidealities which have a substantial impact on detection performance, especially at very low \acp{SNR} \cite{MarKanGio:J15,MarGioChi:J15wcom}: colored noise, noise power uncertainty, in-phase and quadrature-phase imbalance, nonlinearities, spurs, phase noise, and aliasing \cite{6172248,5493999,5506659}. It is therefore essential to perform a proper characterization of the receiver front-end and to design robust detection strategies. 

Regarding noise, it is common to adopt the ubiquitous \ac{AWGN} model \cite{Pro:01,MarGioChi:J11,WeiTirLia:14}. However, wideband receivers are often affected by a non-white noise \ac{PSD}, which corresponds to noise temporal correlation, mainly caused by filtering in the receiver chain \cite{ZenLia:09}. Signal detection in colored noise is studied in \cite{Wha:95} using classical hypothesis testing approaches, while the effect of noise correlation on some eigenvalue based algorithms is analyzed in \cite{Sha:13}.

Another impairment is represented by noise power fluctuations (caused e.g., by receiver temperature variations) often referred as noise uncertainty \cite{SonFis:92,Tor:10,MarGioChi:J11}.
This issue can be counteracted either by adopting noise power estimation strategies \cite{MarGioChi:J11,MarGioChi:15b,MarKanGio:J15} or by adopting detection metrics which are independent on the noise power. In the latter case, time-domain tests that infer the presence of correlation among the received samples have been proposed, in particular exploiting eigenvalues of the \ac{SCM} \cite{ZenLia:09,MarGioChi:12,Sha:13,WeiTirLia:14,MarGioChi:BC14}. 
This is also the case of sensing in the spatial domain, where the correlation among multiple antennas or sensor nodes (by cooperation), due to the presence of a common signal, is exploited \cite{KanGio:12,ZenLia:09,Pen:09,MarGioChi:C15}. 

A key element of \ac{WSS} is that the receiving bandwidth is larger than that of the \ac{StD}, hence for digital receivers oversampling is often the rule \cite{GraMar:74,MarGioChi:J15sp,CaoPes:12}. This fact can be exploited both in the frequency-domain as well as in the time-domain. For example, since oversampling implies correlation in time, detectors based on correlation properties of the received signal are the most common \cite{Gia:98,ZenLia:09,Pen:09,Sha:13,Han:15,Lun:10}.
Another significant element is that the literature on oversampling-based detection explores the case where the \ac{StD} has a known band, i.e., center frequency and bandwidth. However, in some contexts like \acl{OSA}, it is necessary to infer the presence of any signal in the observed frequency slot without knowing its band a priori \cite{MarGioChi:BC14,MarGioChi:15b}.

In this paper, we focus on oversampling-based signal detection in the presence of noise power uncertainty, considering both white and colored Gaussian noise.
The noise model is derived from the statistical analysis of a popular \ac{SDR}, suitable for low-cost, large-scale \ac{RF} monitoring.

The key contributions of this paper are as follows:
\begin{itemize}
\item We analyze the problem of sensing the spectrum in a setting that combines the presence of colored Gaussian noise and noise power uncertainty when the sampling frequency is higher than the signal bandwidth. 
\item We revisit several detectors proposed in the literature, and we derive their performance in terms of \ac{ROC}.
\item We propose new detectors which outperform the existing ones in the specific setting, exhibiting robustness against front-end impairments: a frequency domain version of the \ac{ED} with \ac{ENP}, a detector that measures the spectral flatness, and a spectrum correlation -based detector.
\item Different scenarios are considered: i) unknown signal band; ii) known signal bandwidth but unknown center frequency; iii) both signal bandwidth and center frequency known.
\item The performance is derived in the presence of a signal model extracted from the statistical characterization of the noise of a low-cost \ac{SDR} receiver.
\end{itemize}
The analysis showed that the best detection performance is provided by a noise-uncertainty immune \ac{ED} and, for the colored noise case, by tests that match the \ac{PSD} of the receiver noise.
To the best of our knowledge, \ac{WSS} in this setting is underexplored.

The paper is organized as follows. In Section~\ref{sec:systemmodel} the system model is introduced. Section~\ref{sec:white} proposes detectors in the presence of white noise, while Section~\ref{sec:colored} analyzes detection with colored noise. A case study based on the characterization of a real \ac{SDR} receiver is illustrated in Section~\ref{sec:casestudy}, and the corresponding numerical results are presented in Section~\ref{sec:numericalresults}. Conclusions are drawn in the last section.

Throughout the paper, boldface letters denote matrices and vectors. Moreover, $\Id{m}$ represents the $m\times m$ identity matrix, $\tr{\mathbf{A}}$ is the trace of the matrix $\mathbf{A}$, $\diag{\mathbf{A}}$ stands for a matrix which contains only the principal diagonal of the matrix $\mathbf{A}$, $\transp{(\cdot)}$ and $\hmt{(\cdot)}$ stand, respectively, for simple and Hermitian
transposition.
The $\ell_p$-norm of the vector $\mathbf{v}$ is $\norm{\mathbf{v}}{p}\triangleq\sqrt[p]{\sum_i \left| v_i\right|^p}$, where $v_i$ is the $i$th element of $\mathbf{v}$.

\section{System model}\label{sec:systemmodel}

The detection task is to distinguish the presence or absence of any signal in the observed band. The two hypotheses are denoted, respectively, by $\Ho$ and $\Hz$. 
We consider a receiver equipped with a single antenna.
After down-conversion and sampling, the $\N$-length column vector of the received complex samples is given by
\begin{align}
	\yv{} = \begin{cases}
		\sv{} + \nv{}, & \Ho\\
		\nv{}, & \Hz
	\end{cases}
\end{align}
where $\nv{}$ denotes noise and $\sv{}$ contains the transmitted signal samples including channel effects. Let us consider a bandlimited \ac{StD} with band $\Band=\left[\fL,\fH\right]$ and bandwidth $B=\fH-\fL$.
Oversampling is implemented using a sampling rate $\fs = \OSF \cdot B$, where $\OSF$ is the integer \ac{OSF} \cite{ZenLia:09,Han:15}. 
The \ac{SNR} is defined as $\SNR=\sigs/\sig$, where $\sigs =\EX{\hmt{\sv{}}\sv{}}$ and $\sig=\EX{\hmt{\nv{}}\nv{}}$. 
Without loss of generality we assume that both $\sv{}$ and $\nv{}$ are modeled as vectors of zero mean \acp{r.v.}.

The noise power is often uncertain and varying in time. This is mainly due to effects such as temperature variations, changes in low noise amplifier gain due to thermal fluctuations, and initial calibration errors 
\cite{MarGioChi:J11,Tor:10,SonFis:92}.
We express the noise samples vector as 
$\nv{}=\sigma\,\ntildev{}$, where $\ntildev{}$ are zero mean complex Gaussian samples with unit variance, and the noise power $\sigma^2$ is an unknown time-varying parameter. However, its variations are generally slow, and thus $\sigma^2$ can be considered constant during the collection of the $N$ samples \cite{Tor:10}. 

The most common assumption in the literature is to model noise as a white Gaussian process. This is a desirable condition for every receiver, and it is a realistic model for well-designed systems. However, low-cost wideband devices generally present a colored noise \ac{PSD}, which is mainly due to filtering  \cite{Wha:95,ZenLia:09,Sha:13}. A statistical description of the noise derived from samples captured by a \ac{SDR} device is presented in Section~\ref{sec:noise}. Based on such analysis, we model noise as a correlated Gaussian process.

The detection tests proposed in the next sections exploit the knowledge of the second-order statistical properties of receiver noise, although noise power remains unknown.
We consider, therefore, that detection is preceded by an off-line calibration stage (under hypothesis $\Hz$), in which the system estimates either the noise \ac{PSD}, ${\mathcal W}(f)$, for frequency-domain detectors, or the noise covariance matrix, $\Sig_0$, for time-domain detectors (see Section~\ref{sec:calibration} for details). In both cases, due to noise uncertainty, the noise \ac{PSD} and the covariance matrix are known except for a multiplicative factor related to the time-varying noise power. 
\footnote{Note that although the detectors presented do not depend on the noise power, its fluctuations reflect on the \ac{SNR}, which in turn have an impact on the performance of the tests. This aspect must be taken into account in the study of the decision threshold setting, which is out of the scope of the paper.}
\subsection{Frequency-domain representation: periodogram}

Frequency-domain spectrum sensing is based on the estimation of a spectral representation of received samples and adoption of a test to infer the presence or absence of a signal.
For simplicity, the frequency-domain representation is based on the \ac{PSD} estimation through the averaged periodogram (also known as Bartlett's periodogram), by a $\Nfft$-points \ac{DFT} \cite[Section 12.2.1]{ProMan:96}. In particular, assuming that the total number of samples is $\N = \Nfft \Nsamp$, the $i$th element of the averaged periodogram $\P=(p_0,\dotsc,p_{\Nfft-1})$ is computed as \cite{ProMan:96}
\begin{align} \label{eq:periodogram}
	p_i = \frac{1}{\Nfft \,\Nsamp}\sum_{k=1}^{\Nsamp} \left|\sum_{m=1}^{\Nfft} y_{m+(k-1)\Nfft}\,e^{-j2\pi \frac{i m}{\Nfft}}\right|^2\!
\end{align}
where $y_l$ is the $l$th element of the received vector $\yv{}$.

The Bartlett's periodogram can be used also to estimate the noise \ac{PSD} during the calibration phase. In the following, we denote with $\W=(w_0,\dots,w_{\Nfft-1})$ the vector containing the estimate of the noise \ac{PSD}, ${\mathcal W}(f)$, calculated as
\begin{align} \label{eq:periodogNOISE}
	w_i = \frac{1}{\Nfft \,\Nsamp}\sum_{k=1}^{\Nsamp} \left|\sum_{m=1}^{\Nfft} n_{m+(k-1)\Nfft}\,e^{-j2\pi \frac{i m}{\Nfft}}\right|^2.
\end{align}

\subsection{Time-domain representation: sample covariance}

Using oversampling, time-domain tests can exploit the correlation properties of $\yv{}$ to distinguish the \ac{StD} from \ac{WGN}.
In fact, under $\Hz$ the covariance matrix of white noise is $\Sig_0= \EX{\nv{} \hmt{\nv{}}}=\sig \Id{N}$ and thus its eigenvalues are all equal to $\sig$. Conversely, under $\Ho$, the eigenvalues are not all equal. Therefore, eigenvalue-based tests measure the eigenvalue spread to discriminate between $\Hz$ and $\Ho$. 

When the signal covariance $\Sig= \EX{\yv{} \hmt{\yv{}}}$ is unknown, the \ac{SCM} of $\yv{}$ is used instead \cite{ZenLia:09,MarGioChi:12,Sha:13,WeiTirLia:14}. In this case, the detector arranges the received vector $\yv{}$ in a $\p \times q$ matrix ($p$ and $q$ are such that $N \geq pq$)
\begin{align}\label{eq:Y}
\Y = \begin{bmatrix}
\,\,\, y_1 	 & y_{\p+1} & \dots & y_{(q-1)\p+1}\,\,\, \\
\,\,\, y_2 	 & y_{\p+2} & \dots & y_{(q-1)\p+2}\,\,\, \\
\,\,\,\,\dots 	 & \dots & \dots & \dots \\
\,\,\, y_{\p} & y_{2\p} 	  & \dots & y_{q \p}\,\,\,
\end{bmatrix}.
\end{align}
Then, the eigenvalues of the \ac{SCM} $\SCM = \Y\hmt{\Y}/q$, denoted as $\lambda_{1}\geq\lambda_{2}\geq\dots\geq\lambda_{\p}$, are used as estimate of the eigenvalues of $\Sig$.
Previous works adopt $\p=\OSF$ and $q=\lfloor\N/\p\rfloor$ \cite{Sha:13,Pen:09}. In this case, the rows of $\Y$ are sequences obtained using a sampling period equal to the symbol duration. Assuming the \ac{StD} composed by independent symbols, if $\p=\OSF$ the rows of $\Y$ tend to be independent, while columns are correlated.\footnote{In this case the oversampling-based detection problem turns out to be equivalent to spectrum sensing with multiple antennas.}

Alternatively, some tests are based on the sample correlation matrix obtained by normalizing the \ac{SCM} as
\begin{equation}\label{eq:sampcorr}
\R=\diag{\SCM}^{-1/2} \, \SCM \, \diag{\SCM}^{-1/2}.
\end{equation}
The sample correlation matrix element $r_{ij}$ is the Pearson correlation coefficient between the columns $i$ and $j$ of $\Y$. In the following, we denote with $\mu_{1}\geq\mu_{2}\geq\dots\geq\mu_{\p}$ the eigenvalues of $\R$.

\section{White Noise}\label{sec:white}
In this section, we present \ac{WSS} detectors in the presence of white Gaussian noise and noise power uncertainty.

\subsection{Frequency-domain detectors}\label{sec:FreqDomWhite}

\subsubsection{Energy-based detectors}
The conventional implementation of the \ac{ED} considers the received signal power as a test statistic. Thus, the frequency-domain version of the \ac{ED} is given by\footnote{Note that by Parseval's theorem we have $\TedA=\frac{1}{\Nsamp} \sum_i\hmt{\yvi{}}\yvi{}$, which is proportional to the usual \ac{ED} metric \cite{MarGioChi:J11}.}\footnote{In the paper $\xi$ denotes any detection threshold.}
\begin{align}\label{eq:TedA}
\TedA =  \norm{\P}{1} \testHoneOverHzeroL \xi.
\end{align}
When the \ac{StD} bandwidth is smaller than the receiver bandwidth, the test statistic $\TedA$ \eqref{eq:TedA} also includes the noise-only contributions that come from samples of the \ac{PSD} which are out of the signal band.
It is, therefore, reasonable to modify the \ac{ED} metric, including only the frequency components that may contain the signal. Denoting as $\PB$ the vector that contains the periodogram bins for $\fL\leq f\leq\fH$ 
we propose the test 
\begin{align}\label{eq:Ted}
\Ted = \norm{\PB}{1} \testHoneOverHzeroL \xi.
\end{align}

Note that \eqref{eq:TedA} and \eqref{eq:Ted} depend on the noise power and thus may suffer noise uncertainty. To counteract this problem, schemes that compute the \ac{ENP} can be adopted \cite{MarGioChi:J11}. In the oversampling scenario, the noise power can be estimated from the noise-only bands.
Thus, the frequency-domain version of the \ac{ENP}-\ac{ED} test is given by
\begin{align}\label{eq:Tenped}
\Tenped = \frac{\norm{\PB}{1}}{\norm{\PfsB}{1}} \testHoneOverHzeroL \xi
\end{align}
where $\PfsB$ is the vector containing the periodogram bins that are out of the signal band.
In the Appendix, we prove that $\Tenped$ is the \ac{GLRT} when the \ac{StD} is modeled as a bandlimited Gaussian random process with flat \ac{PSD} within $\Band$.

\subsubsection{Flatness-based detectors}\label{sec:flatness}

In the presence of white noise, some frequency-domain tests are based on the
measure of the \emph{flatness} of the received signal \ac{PSD}. A flat spectrum is expected, indeed, under $\Hz$, contrarily $\Ho$ occurs. 
An example is the \ac{AGM} \cite{GraMar:74} 
\begin{align} \label{eq:Tfagm}
\Tfagm = \frac{\frac{1}{\Nfft}\sum_{i=1}^{\Nfft} p_i}{\left(\prod_{i=1}^{\Nfft}p_i\right)^{1/\Nfft}} \testHoneOverHzeroL \xi.
\end{align}
Another way to measure spectrum flatness is to build a test which is the ratio between the $\ell_1$-norm and the $\ell_2$-norm of $\P$, as 
\begin{align} \label{eq:Tsf}
\Tsf = \frac{1}{\sqrt{\Nfft}}\frac{\norm{\P}{1}}{\norm{\P}{2}} \testHzeroOverHoneL \xi.
\end{align}
The rationale behind \eqref{eq:Tsf} is that its inverse is similar to the population \ac{CV}, a statistical parameter defined as the ratio of the sample standard deviation to the sample mean. The \ac{CV} is often used in statistics and related disciplines as a measure of the variability of a series of samples. In our setting, under $\Hz$ all $w_i$'s are very similar because of the white PSD of noise, so we have appoximatively $\Tsf = 1$ irrespectively of the noise power. Moreover, because of the well-known inequality $\norm{\P}{1}\leq \sqrt{\Nfft}\norm{\P}{2}$ we have $\Tsf\leq 1$.

Note that $\TedA$, $\Tfagm$, and $\Tsf$ do not require the knowledge of $\mathcal B$, which instead must be known for the \ac{ED}-based tests $\Ted$ and $\Tenped$.
Different detectors can be conceived with a partial knowledge of the \ac{StD} characteristics. Considering the case in which only $B$ is known, one can search for the band that contains the maximum received power. Thus, similarly to \eqref{eq:Tenped} we propose the test
\begin{align} \label{eq:TitcB}
\TmosB = \frac{\displaystyle \max_{\BandCB}\norm{\PCB}{1}}{\norm{\P}{1}} \testHoneOverHzeroL \xi
\end{align}
where $\BandCB$ is any frequency slot, with bandwidth $B$, contained in the receiver band.

\subsection{Time-domain detectors}\label{sec:TimeDomWhite}

\subsubsection{Eigenvalue-based tests}\label{sec:timeEigenvalues}

Eigenvalue-based tests are probably the most popular detectors studied in the presence of white noise \cite{ZenLia:09,MarGioChi:12,Sha:13,Pen:09,WeiTirLia:14}.
Of considerable importance there are the sphericity test (also called \ac{AGM}) \cite{Mau:40,MarGioChi:12}, the \ac{MME} \cite{ZenLia:09,Sha:13}, the \ac{MET} \cite{WeiTirLia:14} and the \ac{LBI} test \cite{Joh:71,WeiTirLia:14}. The corresponding decision metrics are given by 
\begin{align}
	&\Tsph = \frac{\left(\prod_{i=1}^{\p}\lambdai\right)^{1/\p}}{ \left(\sum_{i=1}^{\p} \lambdai\right)\!/\p}\testHzeroOverHoneL \xi & \hspace{0.0cm} \Tmme = \frac{\lmax}{\lambda_{\p}}\testHoneOverHzeroL \xi \nonumber\\
	&\Tmet = \frac{\lmax}{\sum_{i=1}^{\p} \lambdai}\testHoneOverHzeroL \xi & \hspace{0.0cm}
	\Tlbi = \frac{\sum_{i=1}^{\p} \lambdai^2}{\left(\sum_{i=1}^{\p} \lambdai\right)^2}\testHoneOverHzeroL \xi. 
\end{align}

We also consider three tests based on the sample correlation matrix $\R$ \eqref{eq:sampcorr}, namely the test of independence \cite{MarGioChi:12}, the Frobenius norm test \cite{LesVdV:01}, and the maximum eigenvalue test, defined, respectively, as 
\begin{align}
&\Tind = \prod_{i=1}^{\p} \mui \testHzeroOverHoneL \xi \hspace{1cm} \Tfro = \sum_{i=1}^{\p} \mui^2\testHoneOverHzeroL \xi \nonumber\\
&\Tmax = \mu_1 \testHoneOverHzeroL \xi.
\end{align}

All the tests defined above are independent of the noise power.
We do not include in the analysis eigenvalue-based tests that suffer noise uncertainty, such as the Wilk's test \cite{WeiTirLia:14}.

\subsubsection{Autocorrelation-based detectors}\label{sec:autocorrelation}

In this section, we discuss some alternative tests that exploit the correlation properties of $\yv{}$, but cannot be expressed as functions of the eigenvalues of the \ac{SCM} or $\R$.
One of the most popular detector in this context is the covariance-based detector \cite{ZenLia:09b}
\begin{align}
\Tcov = \frac{\sum_{n=1}^{\p}\sum_{m=1}^{\p} \left|s_{n,m}\right|}{\tr{\SCM}} \testHoneOverHzeroL \xi.
\end{align}
Some modifications of this detector have been proposed in \cite{Nar:10,Jin:12}.

In \cite{Han:15}, autocorrelation-based approaches specifically for detection in the presence of oversampling have been discussed. In particular, two detectors proposed are 
\begin{align}
	\Tac = \sum_{i=1}^{\OSF-1} \left|\hmt{\yv{\left\{i\right\}}}\,\yv{} + v_i\right|^2 \testHoneOverHzeroL \xi
\end{align}
\begin{align}
	\Tacone = \left|\hmt{\yv{\left\{1\right\}}}\,\yv{} + v_1\right|^2 \testHoneOverHzeroL \xi
\end{align}
where $\yv{\left\{i\right\}}$ denotes a circular shift of $\yv{}$ by $i$ steps, $v_i=\N \alpha_i \sigs /\left( 2 \SNRref+\N \SNRref^2 \left(\alpha_i^2+2\,i^2/\OSF/\N\right)\right)$, and $\alpha_i=\left(\OSF-i\right)/\OSF$. The reference \ac{SNR} parameter $\SNRref$ should be set equal to $\SNR$. However, since in practice, $\SNR$ is unknown, it can be calculated as the \ac{SNR} corresponding to the weakest signal power required to perform detection.

\section{Colored Noise}\label{sec:colored}
In this section, we present \ac{WSS} detectors in the presence of colored Gaussian noise and noise power uncertainty.

\subsection{Frequency-domain detectors}\label{sec:FreqDomColor}

\subsubsection{Energy-based detectors}

Energy-based detectors \eqref{eq:TedA}-\eqref{eq:Tenped} can be used to infer the presence of a signal in the observed band in non-flat noise \ac{PSD} case. 
Regarding \eqref{eq:Tenped}, note that both the numerator and the denominator are proportional to $\sigma^2$; thus, such a detector does not suffer noise uncertainty.

A modification of $\Tenped$ for the colored noise case can be obtained applying \emph{frequency-domain whitening}, consisting of defining the vector ${\bm q}=(q_0,\dots,q_{\Nfft-1})$, with elements $q_i = \frac{p_i}{w_i}$. The resulting test is
\begin{align}\label{eq:Twenped}
\Twenped = \frac{\norm{\QB}{1}}{\norm{\QfsB}{1}} \testHoneOverHzeroL \xi
\end{align}
where $\QB$ and $\QBbar$ are analogous of $\PB$ and $\PBbar$ in \eqref{eq:Tenped}. Note that $w_i$ are the elements of 
$\W$ calculated during the calibration phase by \eqref{eq:periodogNOISE}.

\subsubsection{Flatness-based detectors} 

Considering the \ac{AGM}, after whitening, we get the test \cite{CaoPes:12}
\begin{align} \label{eq:Twfagm}
\Twfagm = \frac{\frac{1}{\Nfft}\sum_{i=1}^{\Nfft} q_i}{\prod_{i=1}^{\Nfft} q_i^{1/\Nfft}} \testHoneOverHzeroL \xi.
\end{align}

\subsubsection{Noise \ac{PSD} matching detectors}

Under hypothesis $\Hz$ the vector $\P$ exhibits a high degree of similarity with $\W$, while they should differ under $\Ho$.
A method to measure the similarity is the degree of correlation between $\P$ and $\W$, which leads to the following test 
\begin{align} \label{eq:Tfc}
\Tfc = \frac{\transp{\P} \W}{\norm{\P}{2} \, \norm{\W}{2}} \testHzeroOverHoneL \xi
\end{align}
where the decision metric is the correlation coefficient between $\P$ and $\W$. 

Alternatively, the \ac{K-L} divergence between $\P$ and $\W$ can be adopted as decision test. In order to avoid noise uncertainty we adopt the normalized vectors $\widetilde{\bm p}={\bm p}/\norm{{\bm p}}{1}$ 
and $\widetilde{\bm w}={\bm w}/\norm{{\bm w}}{1}$, and thus the detector is given by 
\begin{align}\label{eq:fkl}
\Tfkl = \sum_{i=1}^{\Nfft} \widetilde{p}_i\,\log \frac{\widetilde{p}_i}{\widetilde{w}_i} \testHoneOverHzeroL \xi.
\end{align}

Note that both \eqref{eq:Tfc} and \eqref{eq:fkl} are novel. In \cite{So:02} a test similar to \eqref{eq:Tfc} without normalization is adopted when the \ac{StD} spectrum is known a priori and the noise is white. On the contrary, our test assumes that noise is colored, and its \ac{PSD} known from the calibration phase, while the \ac{StD} spectrum is unknown.  

\subsection{Time-domain detectors}\label{sec:TimeDomColor}

When noise is colored, eigenvalue-based algorithms can be adopted if \emph{time-domain whitening} is applied to the received samples before detection \cite{ZenLia:09}.\footnote{Some papers, such as \cite{Sha:13}, propose to use eigenvalue-based tests also in the presence of colored noise, i.e., when the covariance matrix eigenvalues under $\Hz$ are not all equal. 
In this case, the decision regions of the detector and their relative position may depend on the \ac{SNR} and the degree of correlation between noise samples.}
In this domain, whitening is based on the calibration phase described in Section~\ref{sec:systemmodel}, from which the \ac{SCM} of noise under $\Hz$, $\SCMz= \Y\hmt{\Y}/q$, is obtained. Eigenvalue tests can then be applied considering now the eigenvalues $\lambda^{\B}_1\geq\lambda^{\B}_2\geq\dots\geq\lambda^{\B}_\p$ of $\B\, \SCM \, \hmt{\B}$, where $\B$ is the whitening matrix, i.e., a matrix such that $\B\, \SCMz \, \hmt{\B} = \Id{\p}$. Therefore, the sphericity test in the presence of colored noise  becomes 
\begin{align}
\Twsph = \frac{\left(\prod_{i=1}^{\p}\lambdaBi\right)^{1/\p}}{ \left(\sum_{i=1}^{\p} \lambdaBi\right)\!/\p}\testHzeroOverHoneL\xi.
\end{align}
Similarly, considering the tests based on $\R$, we can adopt detectors based on $\mu^{\K}_1\geq\mu^{\K}_2\geq\dots\geq\mu^{\K}_{\p}$, the eigenvalues of $\K \R \hmt{\K}$, where $\K$ is the a matrix such that $\K \Rz \K = \Id{\p}$. For example, the independence test and the maximum eigenvalue test assume the forms
\begin{align}
\Twind= \prod_{i=1}^{\p} \muKi \testHzeroOverHoneL \xi
\end{align}
and
\begin{align}
\Twmax= \mu^{\K}_{1} \testHzeroOverHoneL \xi
\end{align}
respectively.

\begin{table*}[t]
	\centering \begin{tabular}{c||c|c}
		{\it test} & {\it real part} & {\it imaginary part} \\ \hline 
		Jarque-Bera (Univariate Gaussianity) & 0.953 & 0.952 \\ \hline
		Anderson-Darling (Univariate Gaussianity) & 0.965 & 0.966 \\ \hline
		Henze-Zirkler (Multivariate Gaussianity) & 0.941 & 0.942 \\ \hline
		Andersson-Perlman (Circularity) & \multicolumn{2}{c}{0.999}  \\ \hline
	\end{tabular}
	\vspace{0.5cm}
	\caption{Occurrence rate of Gaussianity tests for I-Q samples captured with a RTL-SDR receiver. Univariate tests are based on the observation of $1000$ samples, while multivariate tests consider $4\times 400$ matrices. Each test is performed $50000$ times with a significance level of $0.05$.}
	\label{tab:Gaussianity}
\end{table*}

\section{Case Study: Low-cost SDR Receiver}\label{sec:casestudy}

In this section, we analyze samples collected under $\Hz$ by the RTL-SDR dongle NESDR-mini \cite{noo}. 
Such \ac{SDR} receiver is composed by a Rafael Micro R820T tuner, an analog front-end which includes a low-noise amplifier (with controllable gain via software), \ac{RF} filtering and downconversion at an intermediate frequency of $3.57\,$MHz. The following RTL2832U chip performs an $8\,$bits analog-to-digital conversion, quadrature demodulation, and sample rate reduction through decimation \cite{Hen:99,Skl:16}. The final sampling rate, $\fs$, is adjustable and can be up to $2.8\,$MS/s. I-Q samples are then available on the USB port \cite{realtek:12,Gui:13,Kod:13,MatlabSDR}.

This low-cost device is characterized by the presence of imperfections, such as non-flat noise \ac{PSD}, noise power fluctuations, and spurs caused by harmonics from the mixer, local oscillator leakage and DC offset \cite{CaoPes:12,Bor:15}. This last impairment has not been taken into consideration, assuming that spectrum sensing is preceded by spurious removal, through, e.g., an upstream spur detection and spur censoring stage \cite{CaoPes:12}.

\subsection{Received signal characterization under $\Hz$}\label{sec:noise}

To validate the noise model described in Section~\ref{sec:systemmodel}, we adopt standard statistical Gaussianity tests. 
Table~\ref{tab:Gaussianity} shows the occurrence rate relative to $50000$ tests on captured noise-only samples. All tests are performed with a significance level of $0.05$. 
As univariate tests, we consider the Jarque-Bera and the Anderson-Darling tests, which can be adopted when the mean value and variance are not known \cite{YazYol:07,Ste:74,RazWah:11}. Each test is performed on $1000$ samples.
We also considered a multivariate test, that fits the case in which samples are collected in matrices like \eqref{eq:Y}. 
We chose, in particular, the Henze-Zirkler test proposed in \cite{HenZir:90}, which has an excellent overall power against alternatives to normality \cite{MecMun:03}.
All the tests show that the captured samples well fit Gaussian distributions, with about $0.95$ probability.
Beyond normality, we also tested the circularity property of the samples using the Andersson-Perlman test \cite{AndPer:84,Ada:11}.
These results validate the use of the Gaussian model for the received samples under $\Hz$.

\begin{figure}[t]
\centering
	\psfrag{PSD}{\!\!\!\!\!\!\!\!\!\!\!\!\!\!\!\!\!\!\!\!\!\!\!\!\!\footnotesize $\text{normalized PSD}$ (dB)}
	\psfrag{f}{\!\!\!\!\footnotesize $f/f_\text{s}$}
	\includegraphics[width=0.95\columnwidth,draft=false,clip]{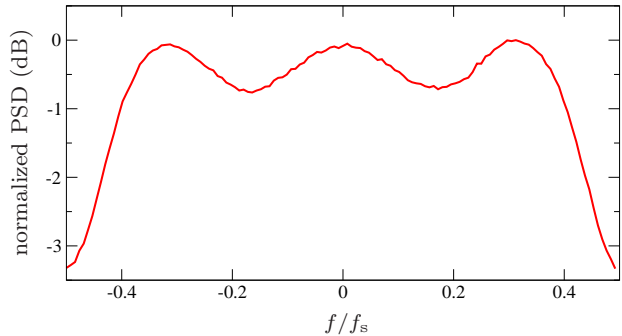}
	\caption{Example of a spurious-free noise \ac{PSD} shape measured in an RTL-SDR dongle. This normalized \ac{PSD} has been estimated from samples, under $\Hz$ hypothesis, using Bartlett's periodogram with $\Nfft=128$ and $\Nsamp=1000$.}
	\label{fig:noise_PSD}
\end{figure}

\subsection{Calibration}\label{sec:calibration}
Regarding the \emph{whiteness} assumption, we estimated the \ac{PSD} in the absence of an input signal. Except for a multiplying constant (i.e., a vertical shift in dB scale), the characteristic shape of the noise \ac{PSD} is shown in Fig.~\ref{fig:noise_PSD}. The multiplying constant is due to the noise power fluctuation, which causes noise power uncertainty, while the shape can be ascribed to the digital filter in the decimation stage of the RTL-SDR receiver chain \cite{Wha:95,ZenLia:09,Sha:13}. 
Based on this analysis, we model noise as a Gaussian process with correlated samples, as described in Section~\ref{sec:systemmodel}.

We assume, therefore, that detection is preceded by an off-line calibration stage in which the system estimates the noise power \ac{PSD} \eqref{eq:periodogNOISE}, for frequency-domain detectors, or the noise covariance matrix, $\Sig_0$, for time-domain detectors. This can be done, e.g., replacing the antenna with a $50\,\Omega$ load. According to the noise model in Section~\ref{sec:systemmodel}, the estimated noise \ac{PSD}, and the \ac{SCM} are known except for a multiplicative constant related to the time-varying noise power.

The accuracy of the estimates performed in the calibration phase typically depends on its duration. A thorough analysis of such a duration is out of the scope of the paper, and it is the subject of future research. As a rule of thumb, the calibration should last at least an order of magnitude more than the detection phase.\footnote{See, for example, \cite{MarGioChi:J11}, in which the length of the noise power estimation phase must be longer than the detection phase, to guarantee that the target $\Pfa$ and $\Pd$ is reached.}

\section{Numerical Results and Discussion}\label{sec:numericalresults}
\subsection{White noise}
The performance of the detectors presented in Section~\ref{sec:white} have been assessed by numerical simulation considering an \ac{OFDM} signal embedded in \ac{AWGN} with $\SNR=-10\,$dB. The bandwidth of the \ac{OFDM} signal equals $\fs / 4$, i.e., $\OSF = 4$.

\subsubsection{Parameters setting}
The performance of frequency-domain detectors introduced in Section~\ref{sec:FreqDomWhite} depends on the total number of samples collected $\N$. Setting $\N$, it is possible to trade-off between $\Nfft$ and $\Nsamp$ for the estimation of the \ac{PSD} in \eqref{eq:periodogram}. In particular, in the following, we set $\N=1600$, and we vary $\Nfft$ and $\Nsamp$ pairs such that $\Nsamp=\lfloor\N/\Nfft \rfloor$.

\begin{figure}[t]
\centering
	\psfrag{Pd}{\scriptsize $\Pd$}
	\psfrag{Pfa}[t][]{\scriptsize $\Pfa$}
	\psfrag{edenp16100}{\scriptsize $\Nfft=16$, $\Nsamp=100$}
	\psfrag{edenp3250}{\scriptsize $\Nfft=32$, $\Nsamp=50$}
	\psfrag{edenp6425}{\scriptsize $\Nfft=64$, $\Nsamp=25$}
	\psfrag{edenp12812}{\scriptsize $\Nfft=128$, $\Nsamp=12$}
	\psfrag{edenp2566}{\scriptsize $\Nfft=256$, $\Nsamp=6$}
	\includegraphics[width=\columnwidth,clip=true]{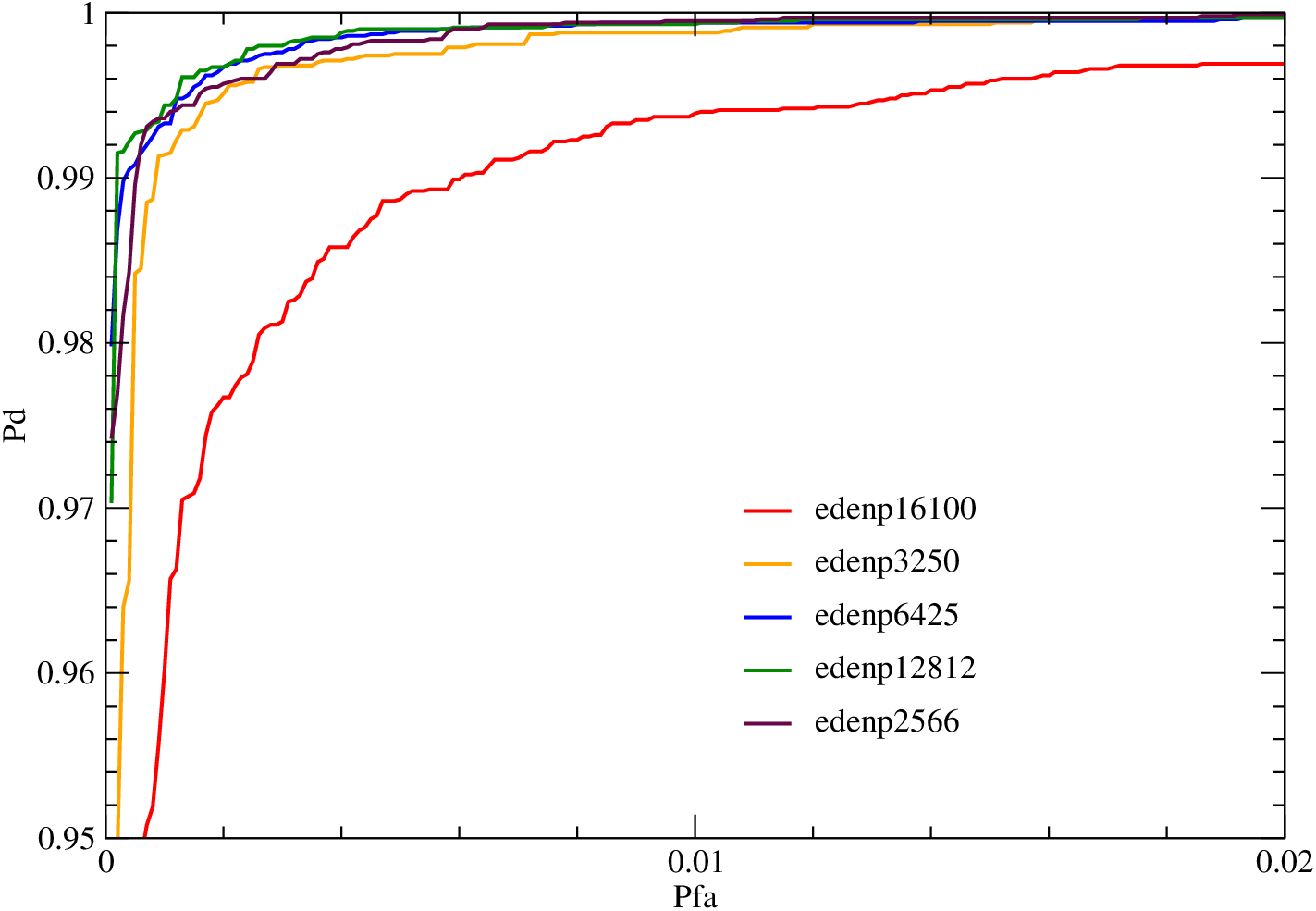}
	\caption{\acp{ROC} for $\Tenped$, varying $\Nfft$ and $\Nsamp$ with $\N=1600$, $\OSF=4$, $\SNR=-10\,$dB.}
	\label{fig:edenp_white}
\end{figure}

Fig.~\ref{fig:edenp_white} shows the \acp{ROC} for $\Tenped$ for different $\Nfft$, $\Nsamp$ pairs. 
The parameter $\Nfft$ impacts the frequency resolution of the \ac{PSD} estimate. A small $\Nfft$ implies, indeed, to have just a few \ac{DFT} elements, each of which collects a large contribution from sidelobes. Therefore, increasing $\Nfft$ provides a better estimate of the in-band signal energy. 
Considering $\Nsamp$, note that it corresponds to the number of periodograms averaged in \eqref{eq:periodogram} and thus it impacts the accuracy of the \ac{PSD} estimate. Therefore, there is a trade-off between these two parameters that provide the best performance. From Fig.~\ref{fig:edenp_white} we can see that the best choice is $\Nfft = 128$ and $\Nsamp=12$. Note that increasing $\Nfft$ above this level, the detector performance decreases due to the small number of averages $\Nsamp$.\footnote{We remark that the optimal value of $\Nsamp$ and $\Nfft$ depends on the specific setting and system parameters.}

\begin{figure}[t]
	\psfrag{Pd}{\scriptsize $\Pd$}
	\psfrag{Pfa}[t][]{\scriptsize $\Pfa$}
	\psfrag{agm16}{\scriptsize $\Tfagm$, $\Nfft=16$, $\Nsamp=100$}
	\psfrag{agm32}{\scriptsize $\Tfagm$, $\Nfft=32$, $\Nsamp=50$}
	\psfrag{agm64}{\scriptsize $\Tfagm$, $\Nfft=64$, $\Nsamp=25$}
	\psfrag{agm128}{\scriptsize $\Tfagm$, $\Nfft=64$, $\Nsamp=25$}
	\psfrag{sf16}{\scriptsize $\Tsf$, $\Nfft=16$, $\Nsamp=100$}
	\psfrag{sf32}{\scriptsize $\Tsf$, $\Nfft=32$, $\Nsamp=50$}
	\psfrag{sf64}{\scriptsize $\Tsf$, $\Nfft=64$, $\Nsamp=25$}
	\psfrag{sf128}{\scriptsize $\Tsf$, $\Nfft=64$, $\Nsamp=25$}
	\centering
	\includegraphics[width=0.985\columnwidth,draft=false,clip]{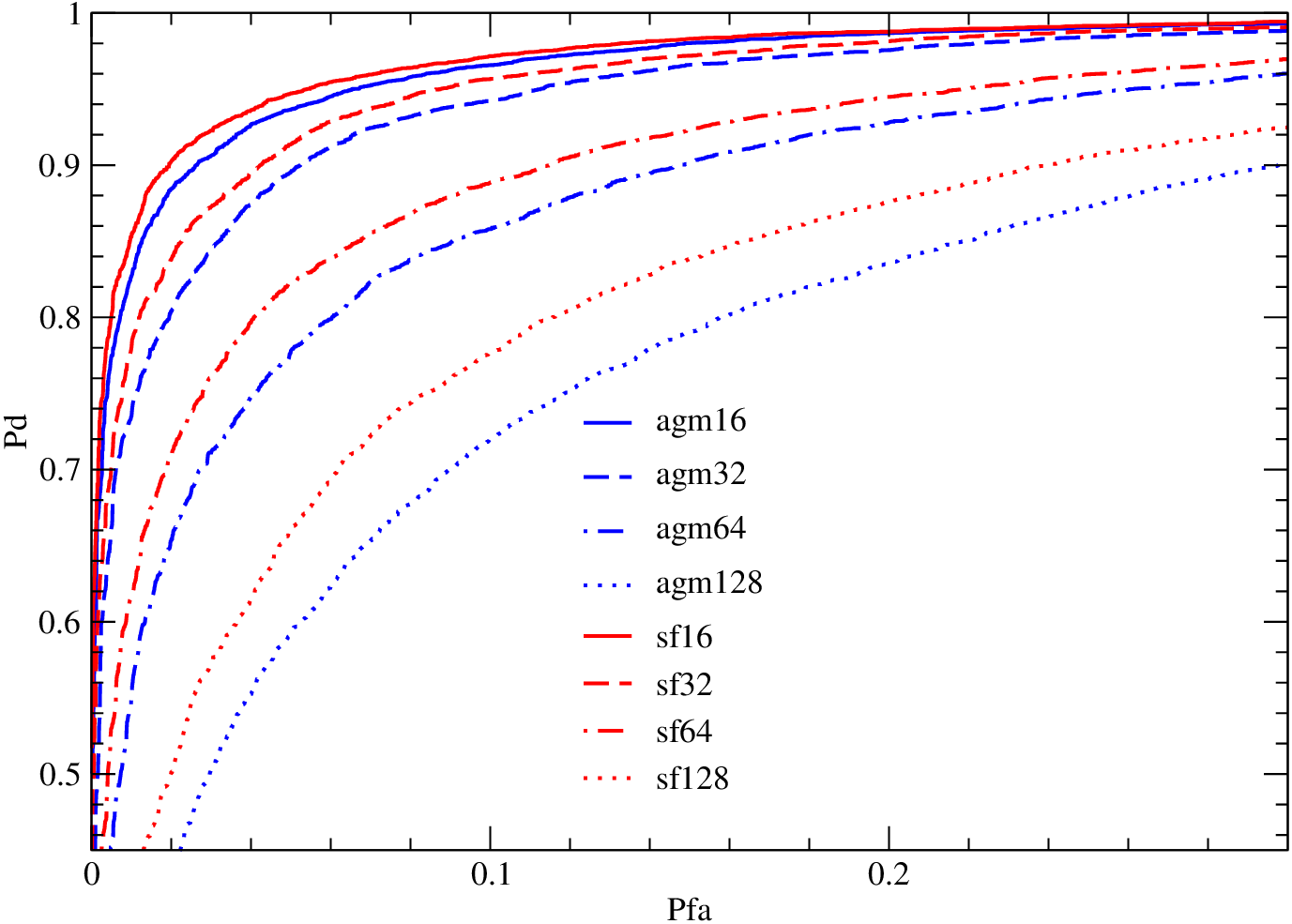}
	\caption{\acp{ROC} for $\Tfagm$ and $\Tsf$, varying $\Nfft$ and $\Nsamp$ with $\N=1600$.}
	\label{fig:freq_white}
\end{figure}

\begin{figure}
	\psfrag{Pd}{\scriptsize $\Pd$}
	\psfrag{Pfa}[t][]{\scriptsize $\Pfa$}
	\psfrag{max4400}{\scriptsize $\Tmax$, $\p=4$, $q=400$}
	\psfrag{max8200}{\scriptsize $\Tmax$, $\p=8$, $q=200$}
	\psfrag{sph4400}{\scriptsize $\Tsph$, $\p=4$, $q=400$}	
	\psfrag{sph 2800}{\scriptsize $\Tsph$, $\p=2$, $q=800$}
	\psfrag{sph8200}{\scriptsize $\Tsph$, $\p=8$, $q=200$}	
	\psfrag{cov4400}{\scriptsize $\Tcov$, $\p=4$, $q=400$}
	\psfrag{cov8200}{\scriptsize $\Tcov$, $\p=8$, $q=200$}
	\psfrag{max cov 2800}{\scriptsize $\Tmax$, $\Tcov$, $\p=2$, $q=800$}
	\centering
	\includegraphics[width=0.97\columnwidth,draft=false,clip]{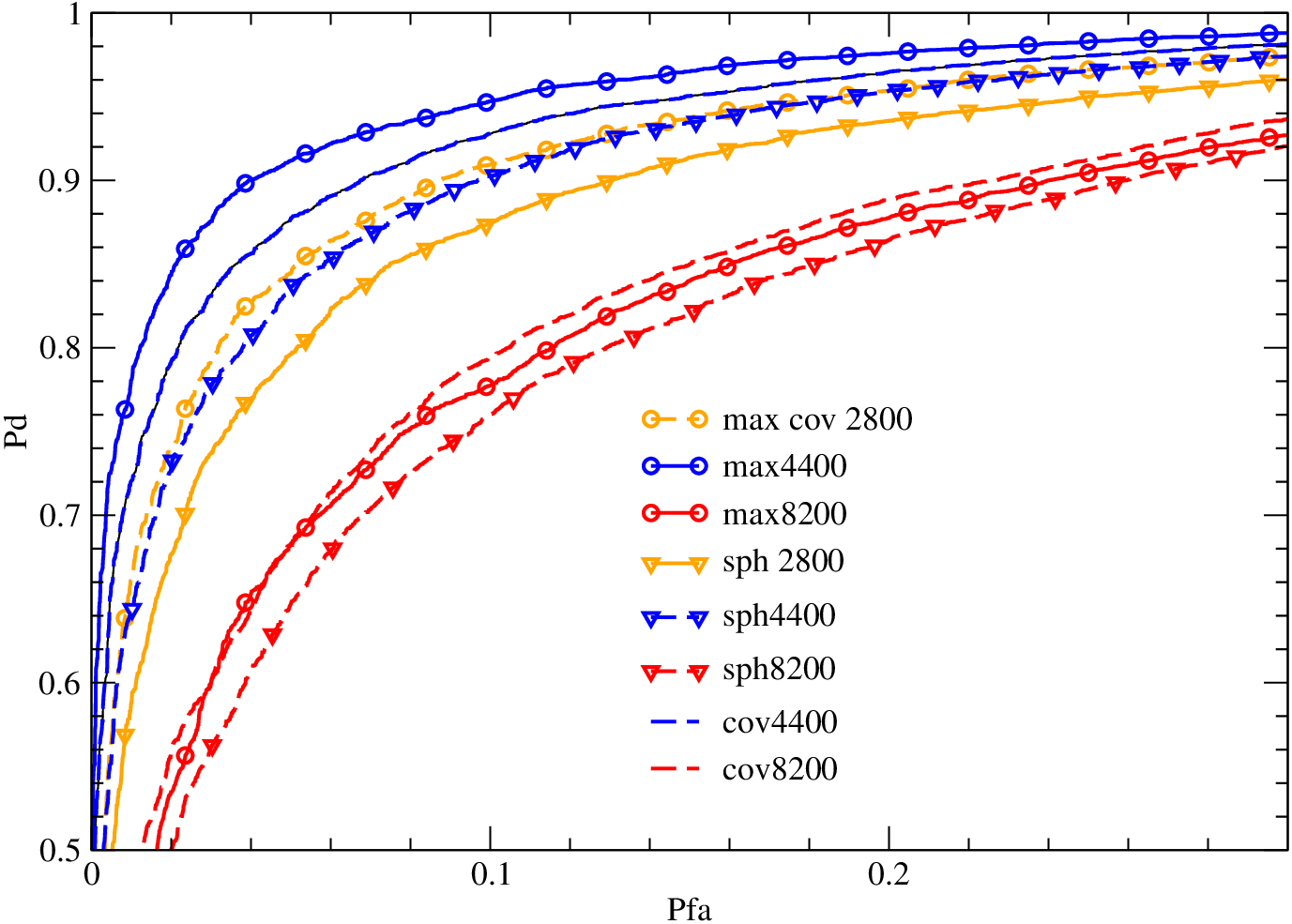}
	\caption{\acp{ROC} for $\Tmax$, $\Tsph$, and $\Tcov$, varying $p$ and $q$ with $\N=pq=1600$.}
	\label{fig:eig_white_param}
\end{figure}

Fig.~\ref{fig:freq_white} shows the \acp{ROC} for the frequency-domain detectors $\Tfagm$ and $\Tsf$ for different $\Nfft$, $\Nsamp$ pairs. Numerical results reveal that the best performance is obtained with $\Nfft=16$. This suggests that for this kind of detectors having a better \ac{PSD} estimate, which can be obtained with a higher $\Nsamp$, is preferable than increasing the frequency resolution.

The parameter setting for the eigenvalue-based detectors is analyzed in Fig.~\ref{fig:eig_white_param}. We can see that the best performance is obtained with $\p=4$, which equals $\OSF$.

\begin{figure}
	\psfrag{Pd}{\scriptsize $\Pd$}
	\psfrag{Pfa}[t][]{\scriptsize $\Pfa$}
	\psfrag{max}{\scriptsize $\Tmax$}
	\psfrag{ind}{\scriptsize $\Tind$}
	\psfrag{sph}{\scriptsize $\Tsph$}	
	\psfrag{cov}{\scriptsize $\Tcov$}
	\psfrag{met}{\scriptsize $\Tmet$}		
	\psfrag{mme}{\scriptsize $\Tmme$}
	\psfrag{fro}{\scriptsize $\Tfro$}
	\psfrag{lbi}{\scriptsize $\Tlbi$}
	\centering
	\includegraphics[width=0.98\columnwidth,draft=false,clip]{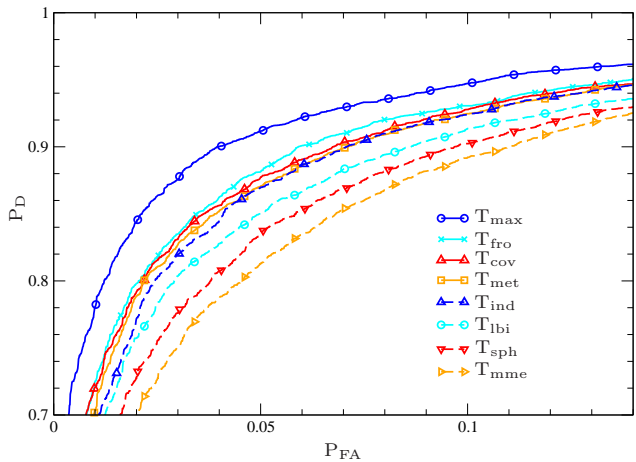}
	\caption{\acp{ROC} for the eivenvalue-based test defined in Section~\ref{sec:TimeDomWhite}, with $\p=4$ and $q=400$.}
	\label{fig:eig_white_248}
\end{figure}

\subsubsection{Detection performance comparison}

Fig.~\ref{fig:eig_white_248} shows the comparison between the eigenvalue-based tests described in Section~\ref{sec:TimeDomWhite}. The best performance is reached by the test $\Tmax$.

\begin{figure}
	\psfrag{Pd}{\scriptsize $\Pd$}
	\psfrag{Pfa}[t][]{\scriptsize $\Pfa$}
	\psfrag{edenp12812}{\scriptsize $\Tenped$, $\!\Nfft=128$, $\!\Nsamp=12$}
	\psfrag{sf 16100}{\scriptsize $\Tsf$, $\!\Nfft=16$, $\!\Nsamp=100$}
	\psfrag{agm 16100}{\scriptsize $\Tfagm$, $\!\Nfft=16$, $\!\Nsamp=100$}
	\psfrag{itc 16100}{\scriptsize $\Tmos$, $\!\Nfft=16$, $\!\Nsamp=100$}
	\psfrag{itcB3 16100}{\scriptsize $\TmosB$, $\!\Nfft=16$, $\!\Nsamp=100$}
	\psfrag{max 4400}{\scriptsize $\Tmax$, $\p=4$, $q=400$}
	\psfrag{corr1-10dB}{\scriptsize $\Tacone$, $\SNRref=-10\,$dB}
	\psfrag{corr-20dB}{\scriptsize $\Tacone$, $\SNRref=-20\,$dB}
	\psfrag{corrA-10dB}{\scriptsize $\Tac$, $\SNRref=-10\,$dB}
	\psfrag{corrA-20dB}{\scriptsize $\Tac$, $\SNRref=-20\,$dB}
	\centering
	\includegraphics[width=\columnwidth,draft=false,clip]{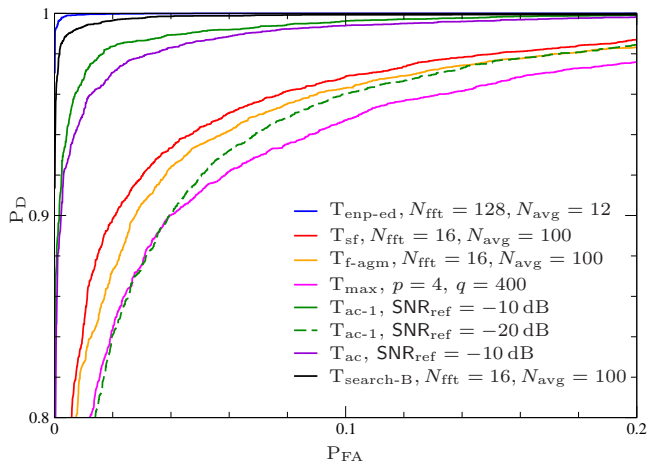}
	\caption{\acp{ROC} for frequency-domain test and time-domain detectors in the white noise scenario, varying $\Nfft$ and $\Nsamp$, with $\N=1600$, $\OSF=4$, and $\SNR=-10\,$dB.}
	\label{fig:white_all}
\end{figure}

The general comparison between all the detectors proposed for the white noise case is presented in Fig.~\ref{fig:white_all}. 
The test $\Tenped$ provides the best detection performance considering both frequency-domain and time-domain tests.
The comparison also include the autocorrelation detectors defined in Section~\ref{sec:autocorrelation}. We can see that $\Tacone$ outperforms $\Tac$ and both provide a good detection performance when $\SNRref=\SNR=-10\,$dB, outperforming all other detectors except for the $\Tenped$. However, $\SNR$ is not known a priori in general, and considering a worst case approach, as proposed in \cite{Han:15}, the detection performance decreases substantially. See, e.g., the case $\SNRref=-20\,$dB. Fig.~\ref{fig:white_all} also reports the comparison between the flatness-based detectors defined in Section~\ref{sec:flatness} and $\Tmax$, which is the best among the eigenvalue-based tests. Note that both $\Tsf$ and $\Tfagm$ outperform $\Tmax$.
From the analysis above we can see that in general, $\Tenped$ is the best detector in the presence of oversampling. When the bandwidth and the carrier frequency of the \ac{StD} are unknown, the flatness-based detectors, and in particular our proposed $\Tsf$, provide a higher $\Pd$ with respect to time-domain detectors.

\subsection{Colored noise}
We now present the \ac{ROC} curves of the detectors described in Section~\ref{sec:colored}, using samples captured from the RTL-SDR dongle.
The RTL-SDR receiver is tuned at $430\,$MHz with sampling frequency $\fs=1\,$MS/s. This frequency band has been chosen for being a signal-free band in our laboratory at the Cesena campus of the University of Bologna.
The \ac{StD} is an \ac{OFDM} waveform transmitting independent symbols and having a $250\,$kHz bandwidth, generated using a \ac{USRP} platform. The transmitter has been located in a non-line-of-sight position, and its transmit power has been tuned to have a \ac{SNR} at the receiver equal to $-10\,$dB.

\begin{figure}[t]
	\psfrag{Pd}{\scriptsize $\Pd$}
	\psfrag{Pfa}[b][]{\scriptsize $\Pfa$}
	\psfrag{3250}{\scriptsize $\Nfft=32$, $\Nsamp=50$}
	\psfrag{6425}{\scriptsize $\Nfft=64$, $\Nsamp=25$}
	\psfrag{12812}{\scriptsize $\Nfft=128$, $\Nsamp=12$}
	\psfrag{2566}{\scriptsize $\Nfft=256$, $\Nsamp=6$}
	\psfrag{5123}{\scriptsize $\Nfft=512$, $\Nsamp=3$}
	\centering
	\includegraphics[width=\columnwidth,draft=false,clip]{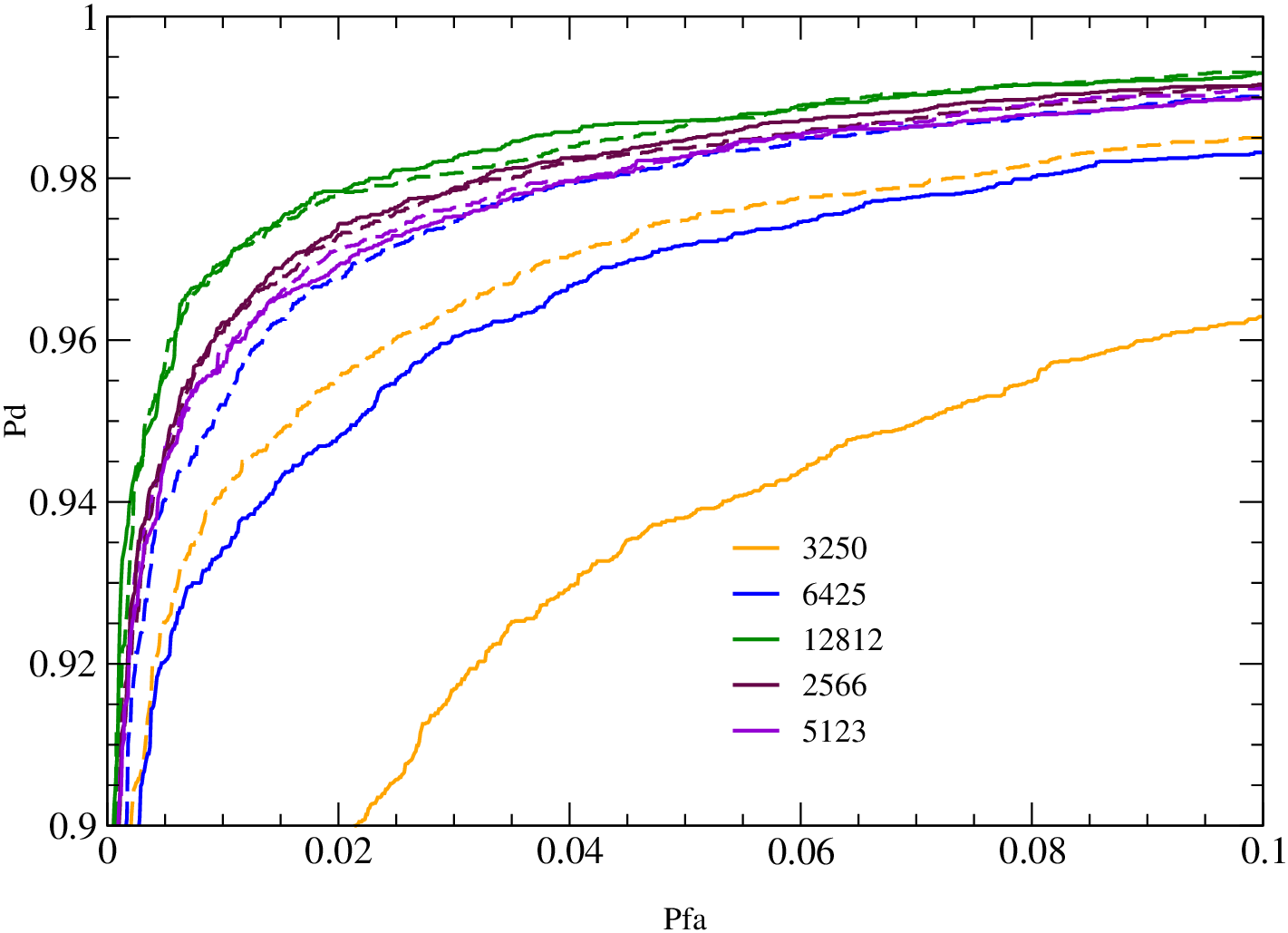}
	\caption{\acp{ROC} for $\Tenped$ (continuous lines) and $\Twenped$ (dashed lines), varying $\Nfft$ and $\Nsamp$ with $\N=1600$. Test on real samples captured with RTL-SDR.}
	\label{fig:edenp_captured}
\end{figure}

\subsubsection{Parameters setting}
In this section, we analyze the trade-off between the parameters $\Nfft$ and $\Nsamp$ for the frequency-based detectors described in Section~\ref{sec:FreqDomColor}.

Fig.~\ref{fig:edenp_captured} shows the \acp{ROC} for $\Tenped$ and $\Twenped$, for different $\Nfft$, $\Nsamp$ pairs. 
From Fig.~\ref{fig:edenp_captured} we can see that also in this case we have a trade-off between $\Nfft$ and $\Nsamp$, and as in the white noise case the best choice is $\Nfft = 128$ and $\Nsamp=12$. 
Regarding the comparison between the whitened and non-whitened \ac{ED}, the plots confirm the advantage introduced by frequency-domain whitening, except for the cases where near-optimal values of $\Nfft$ and $\Nsamp$ are chosen. In these cases, whitening does not provide substantial improvements.

\begin{figure}[t]
	\psfrag{Pd}{\scriptsize $\Pd$}
	\psfrag{Pfa}[t][]{\scriptsize $\Pfa$}
	\psfrag{fc}{\scriptsize $\Tfc$}
	\psfrag{wfagm}{\scriptsize $\Twfagm$}
	\psfrag{fkl}{\scriptsize $\Tfkl$}
	\psfrag{wsf}{\scriptsize $\Twsf$}
	\psfrag{16100}{\scriptsize $\Nfft=16$, $\Nsamp=100$}
	\psfrag{3250}{\scriptsize $\Nfft=32$, $\Nsamp=50$}
	\psfrag{6425}{\scriptsize $\Nfft=64$, $\Nsamp=25$}
	\psfrag{12812}{\scriptsize $\Nfft=128$, $\Nsamp=12$}
	\centering
	\includegraphics[width=0.98\columnwidth,draft=false,clip]{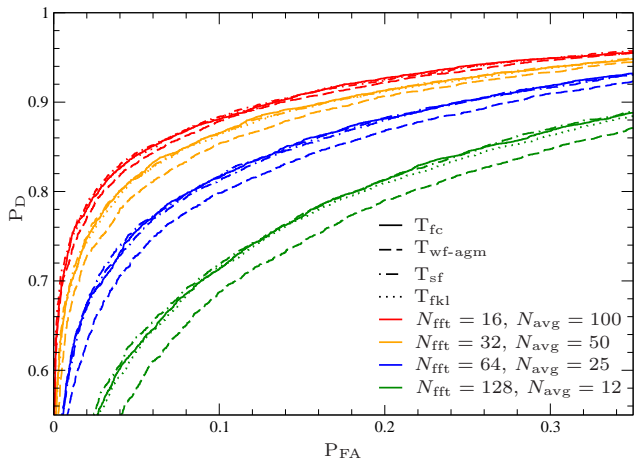}
	\caption{\acp{ROC} for the frequency-based detectors $\Tfc$ (continuous), $\Twfagm$ (dashed), $\Twsf$ (dash-dotted), and $\Tfkl$ (dotted). Test on real samples captured with RTL-SDR.}
	\label{fig:fc_captured}
\end{figure}

In Fig.~\ref{fig:fc_captured}, the \acp{ROC} for $\Tfc$, $\Twfagm$, and $\Tfkl$, are reported for different combinations of $\Nfft$ and $\Nsamp$, respectively. From the comparison, we can see that decreasing $\Nfft$ provides a higher probability of detection.

\subsubsection{Detection performance comparison}
\begin{figure}[t!]
	\psfrag{Pd}{\scriptsize $\Pd$}
	\psfrag{Pfa}[t][]{\scriptsize $\Pfa$}
	\psfrag{edenp128}{\scriptsize $\Tenped$, $\Nfft=128$, $\Nsamp=12$}
	\psfrag{itc16}{\scriptsize $\Tmos$, $\Nfft=16$, $\Nsamp=100$}
	\psfrag{itcB316}{\scriptsize $\TmosB$, $\Nfft=16$, $\Nsamp=100$}
	\psfrag{fc16}{\scriptsize $\Tfc$, $\Nfft=16$, $\Nsamp=100$}
	\psfrag{fwagm16}{\scriptsize $\Twfagm$ $\Nfft=16$, $\Nsamp=100$}
	\psfrag{fkl16}{\scriptsize $\Tfkl$, $\Nfft=16$, $\Nsamp=100$}
	\psfrag{detR4}{\scriptsize $\Twind$, $\p=4$, $\n=400$}
	\psfrag{lbi4}{\scriptsize $\Twlbi$, $\p=4$, $\n=400$}
	\psfrag{sph4}{\scriptsize $\Twsph$, $\p=4$, $\n=400$}
	\psfrag{maxR4}{\scriptsize $\Twmax$, $\p=4$, $\n=400$}
	\centering
	\includegraphics[width=\columnwidth,draft=false,clip]{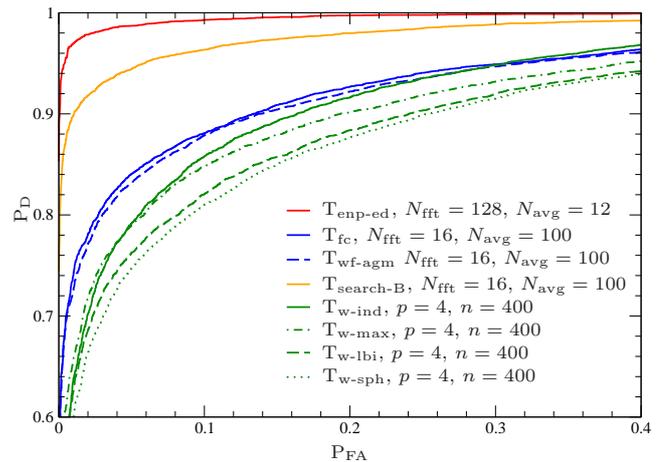}
	\caption{\ac{ROC} comparison of frequency-domain tests and time-domain detectors. Samples captured with RTL-SDR.}
	\label{fig:comparison}
\end{figure}

Fig.~\ref{fig:comparison} shows a comparison between the \acp{ROC} of the  frequency-based detectors of Section~\ref{sec:FreqDomColor} and the time-domain tests described in Section~\ref{sec:TimeDomColor}. We adopt, in this case, the parameters $\Nfft$ and $\Nsamp$ that maximize the detection performance for each test. For the eigevalue-based detectors we adopt $\p=\OSF=4$.
From the comparison, we can see that frequency-domain detectors outperform eigenvalue-based tests. Note, in particular, that $\Tenped$ provides the best performance with much higher detection probability compared to other detectors. For example, when $\Pfa=0.01$  the probability of detection of $\Tenped$ is approximately $\Pd=0.97$, while for $\Tfc$ is $\Pd=0.74$ and for $\Twmax$ is $\Pd=0.65$.

The significant detection performance gain of the \ac{ENP}-\ac{ED} can be explained because $\Tenped$ exploits additional information with respect to other detectors. Note, indeed, that $\Tenped$ requires the knowledge of the signal band, $\Band$. This is a valid assumption, for example, when the signal to be detected is a \ac{PU}, whose channelization is generally known from standards and regulations.

In different \ac{CR} scenarios, however, $\Band$ is unknown, and the best choice is to adopt $\Tfc$, which does not require any knowledge of the signal to be detected (including its operating band) and that provides better detection performance than $\Twfagm$ and eigenvalue-based tests.

\section{Conclusion}\label{sec:conclusions}
In this paper, we discussed the problem of \ac{WSS} in the presence of oversampling, i.e., when the sampling frequency is larger than the signal bandwidth. We considered both the white noise case, which fits well-designed receivers, and the colored noise case, which fits inexpensive devices in which front-end non-idealities cannot be neglected. Considering the latter case, we studied the receiver characteristic of a low-cost commercial device.
We analyzed several detectors, proposing some frequency-domain tests which have been demonstrated to outperform time-domain approaches, such as the standard eigenvalue-based tests.
The analysis showed that the best detection performance is provided by a noise-uncertainty immune \ac{ED} and, for the colored noise case, by tests that match the \ac{PSD} of the receiver noise.


\section*{Appendix}\label{sec:appendix}

The \ac{ENP}-\ac{ED} detector \eqref{eq:Tenped} can be derived as a \ac{GLRT} assuming \ac{AWGN} and a signal described as a white Gaussian process with bandwidth $B$.
Let us organize $\yv{}$ as a sequence of $\Nsamp$ vectors $\yvk$ of length $\Nfft$ with $k=1,\dots,\Nsamp$, and let $\ytildevk$ be the \ac{DFT} of $\yvk$. 
Under the hypotheses made, the spectral representation $\ytildevk$ is Gaussian distributed with covariance $\Sigj$ with $j=0, 1$, depending on the presence/absence of the \ac{StD}. 
Under $\Hz$ its covariance matrix is $\Sigz=\sig \Id{\Nfft}$, while under $\Ho$ the covariance 
$\Sigo$ is a diagonal matrix in which the elements within the signal band are all equal to $\sigt=\sigs+\sig$, and the others equal to $\sig$. 
Thus the joint distribution of the vectors $\ytildevk$ is given by 
\begin{align*}
f\!\left(    \{\ytildevk\}_{k=1}^{\Nsamp} | \Hj   \right)
=&\frac{1}{\pi^{\Nfft \Nsamp}\left|\Sigj\right|^{\Nsamp}} \nonumber\\
& \, \times	\exp\!\left(-\sum_k \hmt{\ytildevk} \Sigj^{-1} \ytildevk\right).
\end{align*}
The \ac{LR} of $\ytildevk$ is therefore
\begin{align} \label{eq:LR}
\mathcal{L} &= { f\!\left(    \{\ytildevk\}_{k=1}^{\Nsamp} | \Ho   \right)}/{f\!\left(    \{\ytildevk\}_{k=1}^{\Nsamp} | \Hz   \right)}\nonumber\\
&=\frac{\left|\Sigz\right|}{\left|\Sigo\right|}\,e^{\Nfft\,\tr{\Sigz^{-1}\SCMf}-\Nfft\,\tr{\Sigo^{-1}\SCMf}}\nonumber\\
&=\frac{\left(\sig\right)^{\Nfft\Nsamp}}{\left(\sig\right)^{\left(\Nfft-|\Sset|\right)\Nsamp}\left(\sigt\right)^{|\Sset|\Nsamp}}\,e^{\Nfft\left((\sig)^{-1}\tr{\SCMf}\right)}\nonumber\\
&\quad\quad\times e^{-\Nfft\left((\sigt)^{-1}\sum_{i\in\Sset}\widetilde{s}_{i,i} + (\sig)^{-1}\sum_{i\in\Ssetbar}\widetilde{s}_{i,i}\right)}
\end{align}
where $\Sset$ is the set of the indexes of the frequency bins within the signal band, $|\Sset|$ is its cardinality and the \ac{SCM} is given by $\SCMf=\sum_{k=1}^{\Nsamp}\, \ytildevk \hmt{\ytildevk}/\Nsamp$.
Under $\Hz$, the \ac{ML} estimate of $\sig$ is $\sigestHz=\sum_{i} s_{i,i}/\Nfft$, while under $\Ho$ we get $\sigestHo=\sum_{i\in\Sset}\stildeii/|\Sset|$ and $\sigestHz=\sum_{i\in\Ssetbar}\stildeii/\left(\Nfft-|\Sset|\right)$. Substituting the estimates into \eqref{eq:LR} we obtain 
\begin{align}\label{eq:LR2}
\mathcal{L} =\kappa\left(1+\frac{\sum_{i\in\Sset}\stildeii}{\sum_{i\in\Ssetbar}\stildeii}\right)^{\!\Nfft\Nsamp} \left(\frac{\sum_{i\in\Sset}\stildeii}{\sum_{i\in\Ssetbar}\stildeii}\right)^{\!-|\Sset|\Nsamp}
\end{align}
where the constant $\kappa$ contains terms that do not depend on the signal samples.
Finally, from \eqref{eq:LR2} we get the sufficient statistic
\begin{align}\label{eq:enped_suff}
\frac{\sum_{i\in\Sset}\stildeii}{\sum_{i\in\Ssetbar}\stildeii}=\frac{\norm{\PB}{1}}{\norm{\PBbar}{1}}
\end{align}
which corresponds to the \ac{ENP}-\ac{ED} test $\Tenped$ in \eqref{eq:Tenped}.

\bibliographystyle{IEEEtran}
\bibliography{IEEEabrv,SDRsensing,bibAM}

\bigskip

\small

\begin{wrapfigure}{l}{0.3\columnwidth}
\centering
\includegraphics[width=0.32\columnwidth]{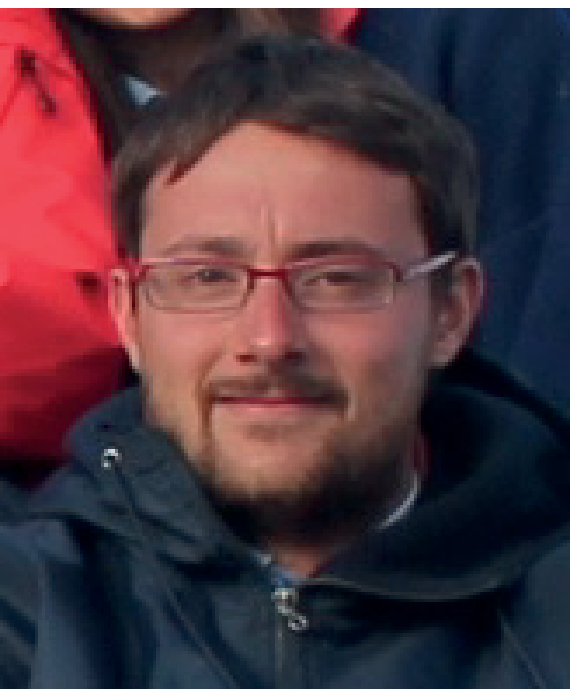}
\end{wrapfigure}
\noindent {\bf Andrea~Mariani} received the B.Sc. degree in Biomedical Engineering (with honors) in 2006, the M.Sc. degree in Electronics and Telecommunications Engineering (with honors) in 2009, and the Ph.D. degree in Electronics, Computer Science and Telecommunications in 2013 from the University of Bologna, Italy. From 2013 to 2016, he was is with the Center for Industrial Research on ICT (CIRI-ICT) of the University of Bologna. In 2016 he joined Onit Group, Italy, as R\&D Manager for the Industry Division. He has participated in several international projects, such as FP7-ICT EUWB, FP7-ICT SELECT and EDA-B CORASMA projects. 

His research interests include statistical signal processing and digital communications, with particular emphasis on localization, spectrum sensing for cognitive radio networks and software defined radio. 

Dr. Mariani was a member of the Technical Program Committee of the Cognitive Radio and Networks Symposium at the IEEE Int. Conf. on Commun. (ICC 2013 - 2016), the Cognitive Radio and Networks Symposium at the IEEE Global Commun. Conf. (Globecom 2013), and was a member of the local organization committee for the IEEE Int. Conf. on Ultra-WideBand (ICUWB 2011). He serves as reviewer for several IEEE Journals and Conferences, and has been named an Exemplary Reviewer for the \textsc{IEEE Wireless Communications Letters} in 2013 and for the \textsc{IEEE Communications Letters} in 2013 and 2014. 

\bigskip
\begin{wrapfigure}{l}{0.3\columnwidth}
\centering
\includegraphics[width=0.32\columnwidth]{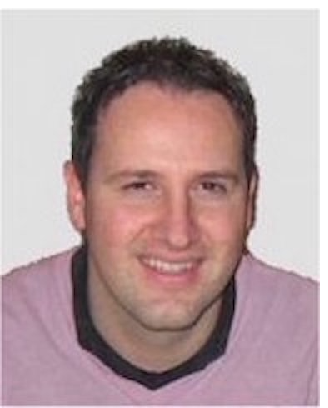}
\end{wrapfigure}
\noindent {\bf Andrea~Giorgetti} received the Dr. Ing. degree (\textit{summa cum laude}) in electronic
engineering and the Ph.D. degree in electronic engineering and computer science from the University of Bologna, Italy, in 1999 and 2003, respectively.
From 2003 to 2005, he was a Researcher with the National Research Council, Italy. He joined the Department of Electrical, Electronic, and Information Engineering ``Guglielmo Marconi,'' University of Bologna, as an Assistant Professor in 2006 and was promoted to Associate Professor in 2014. 
In spring 2006, he was with the Laboratory for Information and Decision Systems (LIDS), Massachusetts Institute of Technology (MIT), Cambridge, MA, USA. Since then, he has
been a frequent visitor to the Wireless Information and Network Sciences Laboratory at the MIT, where he presently holds the Research Affiliate appointment.

His research interests include ultrawide bandwidth communication systems, active and passive localization, wireless sensor networks, and cognitive radio. He has co-authored the book \textit{Cognitive Radio Techniques: Spectrum Sensing, Interference Mitigation, and Localization} (Artech House, 2012).
He was the Technical Program Co-Chair of several symposia at the IEEE Int. Conf. on Commun. (ICC), and IEEE Global Commun. Conf. (Globecom). 
He has been an Editor for the \textsc{IEEE Communications Letters} and for
the \textsc{IEEE Transactions on Wireless Communications}.
He has been elected Chair of the IEEE Communications Society's Radio Communications Technical Committee. 

\bigskip
\begin{wrapfigure}{l}{0.3\columnwidth}
\centering
\includegraphics[width=0.32\columnwidth]{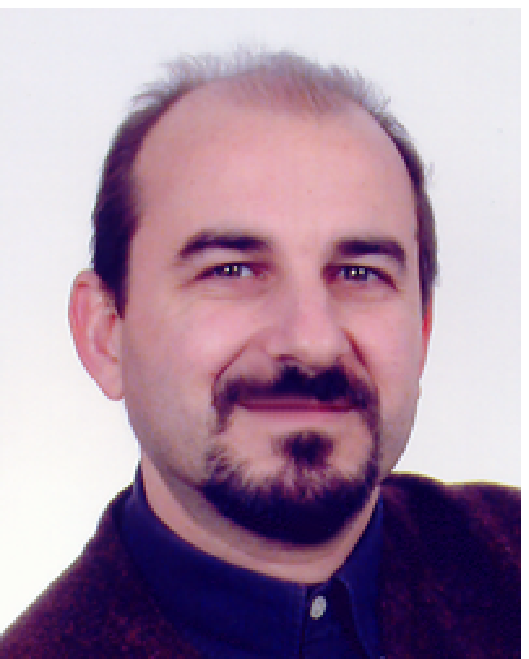}
\end{wrapfigure}
\noindent {\bf Marco~Chiani} received the Dr. Ing. degree (\textit{summa cum laude}) in electronic engineering and the Ph.D. degree in electronic and computer engineering from the University of Bologna, Italy, in 1989 and 1993, respectively.

He is a Full Professor in Telecommunications at the University of Bologna. During summer 2001, he was a Visiting Scientist at AT\&T Research Laboratories, Middletown, NJ. Since 2003 he has been a frequent visitor at the Massachusetts Institute of Technology (MIT), Cambridge, where he presently holds a Research Affiliate appointment.
 His research interests are in the areas of communications theory, wireless systems, and statistical signal processing, including MIMO statistical analysis, codes on graphs, wireless multimedia, cognitive radio techniques, and ultra-wideband radios.

In 2012 he has been appointed Distinguished Visiting Fellow of the Royal Academy of Engineering, UK. He is the past chair (2002--2004) of the Radio Communications Committee of the IEEE Communication Society and past Editor of Wireless Communication (2000--2007) for the journal \textsc{IEEE Transactions on Communications}.
He received the 2011 IEEE Communications Society Leonard G. Abraham Prize in the Field of Communications Systems, the 2012 IEEE Communications Society Fred W. Ellersick Prize, and the 2012 IEEE Communications Society Stephen O. Rice Prize in the Field of Communications Theory.

\end{document}